\documentclass[twocolumn,showpacs,preprintnumbers,amsmath,amssymb]{revtex4}
\usepackage{amsfonts}

\usepackage{graphicx}
\usepackage{dcolumn}
\usepackage{bm}
\usepackage{mathrsfs}
\usepackage{graphicx}
\usepackage{float}
\usepackage{amsmath}
\usepackage{subfigure}

\begin{document}

\title{ Probabilistic chain teleportation of a qutrit-state  }

\author{Meiyu Wang, Fengli Yan}

\email{flyan@mail.hebtu.edu.cn}

\affiliation {~\\
    College of Physics Science and Information Engineering, Hebei Normal University, Shijiazhuang 050016, China\\
   ~
Hebei Advanced Thin Films Laboratory,  Shijiazhuang 050016, China}

\date{\today}

\begin{abstract}
{We investigate chain teleportation  of a qutrit-state via the
non-maximally two-qutrit entangled channels.  For the case of four
parties, the efficiencies of two chain teleportation protocols, the
separate chain teleportation protocol (SCTP) and the global chain
teleportation protocol (GCTP), are calculated. In SCTP the errors
are corrected between every step while in GCTP the errors are
corrected only at the end. Furthermore, we present a piecewise
global chain teleportation protocol (PGCTP) for keeping away from
the inconvenience of error-correction of GCTP.  We show that PGCTP
is more efficient than SCTP.
 }
\end{abstract}

\maketitle

Quantum teleportation plays an important role
   in the field of quantum computation and quantum communication. It allows a sender(called Alice) to transmit an unknown quantum
   state to a receiver(called Bob) by using an entangled state as
   a quantum resource and by sending classical information via ordinary
channel. The original protocol of Bennett et al. \cite{bbcjpw}
involves teleportation of an arbitrary state of a qubit via an
Einstein- Podolsky-Rosen (EPR) pair and by transmitting two bits of
classical information from Alice to Bob. Here Alice knows neither
the state to be teleported nor the location of the intended
receiver, Bob. They also presented a protocol for teleporting an
unknown state of a qudit via a maximally entangled state in $d\times
d$ dimensional Hilbert space and by sending $2\log_2 d$ bits of
classical information. Since then, quantum teleportation has got
great development \cite {Braunstein, Wang, KLM, VaidmanPRA1994,
BKprl1998, GRpra,  KB, GaoYan, APpla, PAjob,
DLLZWpra2005,Zhangzhanjun, MG,Rigolin, WangYan} and has been experimentally
demonstrated by several groups \cite{BPMEWZ,FSBFKP,NKL,Boschi}.

However in the real world situations, it is most of the time not
   possible to have a maximally entangled state at one's disposal.
Because of the interaction with the environment, the state of any
system
   would become a mixed state after a certain period.
This problem of decoherence can be mitigated
   but cannot be completely overcome easily.
Also, it may happen that the source does not produce perfect EPR
pairs rather non-maximally entangled pairs which is shared between
Alice and Bob. If we have non-maximally entangled state as a shared
resource and we want to do quantum teleportation, we have to pay
some price. That is, we have to compromise either in fidelity or in
the success probability. If we are ready to pay the price for the
success probability it is possible to have unit fidelity
teleportation. And this scheme is called probabilistic quantum
teleportation \cite{APpla, PAjob}.

When one considers teleportation in a  quantum chain network, the
teleportation strategy is miscellaneous. Suppose that the quantum
chain network consists of $N+1$ parties  (called Alice1, Alice2,
Alice3, ..., Alice$N+1$ respectively in sequence) who share the
 entangled sources between the neighboring parties. The
simplest and most direct strategy is the separate chain
teleportation protocol (SCTP) by performing $N$ separate
teleportations, i.e., Alice1 teleports the quantum state to Alice2
via the first teleportation. Then Alice2 teleports it to Alice3 via
the second teleportation and so on. If the entangled sources between
every neighboring parties are the same, the total success
probability equals $p^N$ when a quantum state teleportation from
starting point Alice1 to end point Alice$N+1$, where $p$ is the
success probability of one step. The total success probability of
SCTP is very small for $p<1$ when $N$ is large. A more efficient
strategy is the global chain teleportation protocol (GCTP)\cite{MG}. In an interesting work, Mod{\l}awska and Grudka \cite {MG}
showed that if the qubit is teleported several times via some
non-maximally entangled states, the \textquotedblleft errors"
introduced in the previous teleportations can be corrected by the
\textquotedblleft errors" introduced in the following
teleportations. This effect is called error self-correction. The
total success probability would be enhanced immensely if we correct
all errors in the final receiver. Their strategy was developed in
the framework of the scheme proposed in Ref.\cite {KLM} for linear
optical teleportation. In the recent papers \cite{Rigolin,WangYan}, it has been
shown that this feature of the multiple teleportation of
Ref.\cite{MG} is not restricted to the teleportation scheme stated
in Ref.\cite {KLM}. In Ref. \cite {WangYan}, we present global chain teleportation protocol
based on the general teleportation language of the original proposal
shown in Ref. \cite{bbcjpw}.

In this paper, we would investigate the probabilistic chain
teleportation of qutrit-state via non-maximally two-qutrit entangled
channels. For the case of four parties, the efficiencies of two
chain teleportation protocols, the SCTP and the GCTP, are
calculated. With the increase of the parties number $N$, the
collapsing states will become more and more complex if we apply the
GCTP. For keeping away from this inconvenience, we present the
piecewise global chain teleportation protocol (PGCTP). We show that
PGCTP is more efficient than SCTP.

It is easy to see that the error self-correction only appears in the
case of $3N+1$ parties for chain teleportation of qutrit state.
Therefore, to illustrate our protocol clearly, let us first begin
with the chain teleportation  in the case of four parties.

Alice wants to teleport an unknown quantum state
\begin{equation}\label{2}
    |\psi\rangle=\alpha|0\rangle+\beta|1\rangle+\gamma|2\rangle
\end{equation}
to Bob, where $\alpha, \beta$   and  $\gamma \in \mathbb{C}$ and
$|\alpha|^2+|\beta|^2+|\gamma|^2=1$. There is no direct entangled
resource between Alice and Bob. Fortunately, Alice and Bob can build
the relationship by two intermediaries Charlie and Dave. Alice and
Charlie share a non-maximally two-qutrit entangled resource
\begin{equation}\label{1}
    |\Psi\rangle=a_{0}|00\rangle+a_{1}|11\rangle+a_{2}|22\rangle,
\end{equation}
while  Charlie and Dave, Dave and Bob share the identical resource,
where $a_{0}$, $a_{1}$ and $a_{1}$are real numbers and satisfy
$a_{0}^2+a_{1}^2+a_{2}^2=1$. Without loss of generality, we suppose
$a_{0}\leq a_{1}\leq a_{2}$.

Firstly, we consider the SCTP completed by performing three separate
standard teleportations, i.e., Alice teleports the quantum state
$|\psi\rangle$ to Charlie via the first teleportation. Then Charlie
teleports it to Dave via the second teleportation. Finally, Dave
teleports it to receiver Bob via the third teleportation. According
to the standard probabilistic teleportation protocol, in the first
separate teleportation, Alice performs generalized Bell-basis
measurement(GBM) in the basis $\{|\Phi_{mn}\rangle, m, n=0, 1,
2\}$on the teleported qutrit and the entangled qutrit in her side.
\begin{equation}\label{GBM}
|\Phi_{mn}\rangle=\frac{1}{\sqrt{3}}\sum_{j=0}^{2}e^{2ijn\pi/3}|j\rangle|(j\oplus
m)\rangle,
\end{equation}
where here and hereafter $j \oplus m$ means sum of $j$ and $m$
modulo $3$.
 Charlie can apply the corresponding unitary
transformation conditioned on the result of Alice's measurement. The
corresponding unitary transformations read
\begin{equation*}
\begin{array}{ll}
U_{00}=\left [\begin{array}{ccc} 1&0&0\\
0&1&0\\
0&0&1\end{array}\right], U_{01}=\left [\begin{array}{ccc} 1&0&0\\
0&e^{-2\pi i/3}&0\\
0&0&e^{-4\pi i/3}\end{array}\right],\\[0.5cm]
U_{02}=\left [\begin{array}{ccc} 1&0&0\\
0&e^{-4\pi i/3}&0\\
0&0&e^{-2\pi i/3}\end{array}\right],\\
U_{10}=\left [\begin{array}{ccc} 0&0&1\\
1&0&0\\
0&1&0\end{array}\right],U_{11}=\left [\begin{array}{ccc} 0&0&e^{-4\pi i/3}\\
1&0&0\\
0&e^{-2\pi i/3}&0\end{array}\right], \\[0.5cm]
U_{12}=\left [\begin{array}{ccc} 0&0&e^{-2\pi i/3}\\
1&0&0\\
0&e^{-4\pi i/3}&0\end{array}\right],\\[0.5cm]
\end{array}\end{equation*}
\begin{equation}
\begin{array}{ll}
U_{20}=\left [\begin{array}{ccc} 0&1&0\\
0&0&1\\
1&0&0\end{array}\right], U_{21}=\left [\begin{array}{ccc} 0&e^{-2\pi i/3}&0\\
0&0&e^{-4\pi i/3}\\
1&0&0\end{array}\right],\\[0.5cm]
U_{22}=\left [\begin{array}{ccc} 0&e^{-4\pi i/3}&0\\
0&0&e^{-2\pi i/3}\\
1&0&0\end{array}\right].\end{array}\end{equation}
Finally, the state
Bob received becomes
\begin{equation}\label{4}
    |\psi_{1}\rangle=\frac{1}{\sqrt{p_1}}(\alpha a_{0}|0\rangle+\beta
    a_{1}|1\rangle+\gamma a_{2}|2\rangle)
\end{equation}
with the probability
$p_1=|a_{0}\alpha|^2+|a_{1}\beta|^2+|a_{2}\gamma|^{2}$ or
\begin{equation}\label{4}
    |\psi_{2}\rangle=\frac{1}{\sqrt{p_3}}(\alpha a_1|0\rangle+\beta a_2|1\rangle+\gamma
   a_0|2\rangle)
\end{equation}
with the probability
$p_2=|a_1\alpha|^2+|a_2\beta|^{2}+|a_0\gamma|^2$ or
\begin{equation}\label{4}
    |\psi_{3}\rangle=\frac{1}{\sqrt{p_2}}(\alpha a_{2}|0\rangle+\beta a_{0}|1\rangle+\gamma
   a_{1}|2\rangle)
\end{equation}
with the probability
$p_3=|a_{2}\alpha|^2|+|a_{0}\beta|^2+|a_{1}\gamma|^2$.

 These states
are in accordance with the original state $|\psi\rangle$ only if the
quantum channel is a maximally entangled state, i.e. $a_0=a_1=a_2$.
For the case of  non-maximally entangled channel, these states can
be returned to the original state with certain probability by
performing the generalized measurerment given by Kraus operators:
\addtocounter{equation}{1}
\begin{align}
E_{S1}&=| 0 \rangle\langle 0 |+\frac{a_0}{a_1}| 1 \rangle\langle 1 |+\frac{a_0}{a_2}|2\rangle\langle2|, \tag{\theequation a}\\
E_{F1}&= \sqrt{1-\frac{a_0^2}{a_1^2}} | 1 \rangle\langle
1|+\sqrt{1-\frac{a_0^2}{a_2^2}} | 2 \rangle\langle 2 |
\tag{\theequation b}
\end{align}
for $|\psi_1\rangle$ and \addtocounter{equation}{1}
\begin{align}
E_{S2}&=\frac{a_0}{a_1}| 0 \rangle\langle 0 |+\frac{a_0}{a_2}|1 \rangle\langle 1|+| 2 \rangle\langle 2 |, \tag{\theequation a}\\
E_{F2}&= \sqrt{1-\frac{a_0^2}{a_1^2}} | 0 \rangle\langle
0|+\sqrt{1-\frac{a_0^2}{a_2^2}} |1\rangle\langle1 |\tag{\theequation
b}
\end{align}
for $|\psi_2\rangle$ and \addtocounter{equation}{1}
\begin{align}
E_{S3}&=\frac{a_0}{a_2}| 0 \rangle\langle 0 |+| 1 \rangle\langle 1 |+\frac{a_0}{a_1}|2 \rangle\langle 2|, \tag{\theequation a}\\
E_{F3}&= \sqrt{1-\frac{a_0^2}{a_2^2}} | 0 \rangle\langle
0|+\sqrt{1-\frac{a_0^2}{a_1^2}} |2\rangle\langle2| \tag{\theequation
b}
\end{align}
for $|\psi_3\rangle$.

 When $E_S$ is obtained, the qutrit ends in its
original state
$|\psi\rangle=\alpha|0\rangle+\beta|1\rangle+\gamma|2\rangle$. The
success probability in the first teleportation is
\begin{equation}\label{6}
    p=\sum_{i=1}^3p_i\langle\psi_i|E^\dag_{Si}E_{Si}|\psi_i\rangle=3a
    _0^2.
\end{equation}
  Next, Charlie teleports the recovered quantum state to Dave, and Dave teleports the recovered quantum state to Bob by the
similar process. Combining these three teleportations, the total
probability that Bob receives the quantum state $|\psi\rangle$ is
\begin{equation}\label{7}
    P_{S}(4)=p^3=27a_0^6.
\end{equation}

However, the above teleportation protocol is not the optimal
strategy. In fact, the intermediaries Charlie and Dave  does not
need to recover the quantum state to be teleported. They only make
the conditioned unitary transformation according to the previous
party's measurement outcome and teleports the \textquotedblleft
errors" state to the next party directly. Lastly, Bob corrects all
\textquotedblleft errors" of the quantum state in the teleportation
process. This is so called GCTP.

Let us, thus, assume that Charlie and Dave does not recover the
quantum state to be teleported in the teleportation, they only make
a unitary transformation according to the previous party's
measurement outcome, then they  also perform GBM on their two
qutrits and broadcast the measurement outcome to the next party.
After making the corresponding transformation conditioned on Dave's
measurement outcome, Bob's qutrit will collapse into one of the
following states

 \addtocounter{equation}{1}
\begin{align}
&|\phi_1\rangle = a_0^3\alpha|0\rangle+ a_1^3\beta|1\rangle+
    a_2^3\gamma|2\rangle, \tag{\theequation a}\\
&|\phi_2\rangle = a_1^3\alpha|0\rangle+ a_2^3\beta|1\rangle+
    a_0^3\gamma|2\rangle, \tag{\theequation b}\\
&|\phi_3\rangle = a_2^3\alpha|0\rangle+ a_0^3\beta|1\rangle+
    a_1^3\gamma|2\rangle, \tag{\theequation c}\\
&|\phi_4\rangle = a_0^2a_1\alpha|0\rangle+
   a_1^2a_2\beta|1\rangle+ a_2^2a_0\gamma|2\rangle,\tag{\theequation d}\\
&|\phi_5\rangle = a_1^2a_2\alpha|0\rangle+
   a_2^2a_0\beta|1\rangle+ a_0^2a_1\gamma|2\rangle,\tag{\theequation e}\\
&|\phi_6\rangle = a_2^2a_0\alpha|0\rangle+
   a_0^2a_1\beta|1\rangle+ a_1^2a_2\gamma|2\rangle,\tag{\theequation f}\\
&|\phi_7\rangle = a_0^2a_2\alpha|0\rangle+
   a_1^2a_0\beta|1\rangle+ a_2^2a_1\gamma|2\rangle,\tag{\theequation g}\\
&|\phi_8\rangle = a_1^2a_0\alpha|0\rangle+
   a_2^2a_1\beta|1\rangle+ a_0^2a_2\gamma|2\rangle,\tag{\theequation h}\\
&|\phi_9\rangle = a_2^2a_1\alpha|0\rangle+
   a_0^2a_2\beta|1\rangle+ a_1^2a_0\gamma|2\rangle,\tag{\theequation i}\\
&|\phi_{10}\rangle =\alpha|0\rangle+\beta |1\rangle+\gamma
   |2\rangle \tag{\theequation j}
\end{align}
with the probabilities
\begin{eqnarray*}
  p^\prime_1 &=&  a_0^6|\alpha|^2+a_1^6|\beta|^2+a_2^6|\gamma|^2, \\
   p^\prime_2 &=&  a_1^6|\alpha|^2+a_2^6|\beta|^2+a_0^6|\gamma|^2, \\
   p^\prime_3 &=&  a_2^6|\alpha|^2+a_0^6|\beta|^2+a_1^6|\gamma|^2, \\
 p^\prime_4 &=& 3( a_0^4a_1^2|\alpha|^2+a_1^4a_2^2|\beta|^2+a_2^4a_0^2|\gamma|^2), \\
p^\prime_5 &=& 3(a_1^4a_2^2|\alpha|^2+a_2^4a_0^2|\beta|^2+a_0^4a_1^2|\gamma|^2), \\
p^\prime_6 &=& 3( a_2^4a_0^2|\alpha|^2+a_0^4a_1^2|\beta|^2+a_1^4a_2^2|\gamma|^2), \\
p^\prime_7 &=& 3( a_0^4a_2^2|\alpha|^2+a_1^4a_0^2|\beta|^2+a_2^4a_1^2|\gamma|^2), \\
p^\prime_8 &=& 3( a_1^4a_0^2|\alpha|^2+a_2^4a_1^2|\beta|^2+a_0^4a_2^2|\gamma|^2), \\
p^\prime_9 &=&  3(a_2^4a_1^2|\alpha|^2+a_0^4a_2^2|\beta|^2+a_1^4a_0^2|\gamma|^2), \\
p^\prime_{10} &=& 6a_0^2a_1^2a_2^2.
 \end{eqnarray*}
respectively. When the state is in $|\phi_{10}\rangle$, we do not
have to perform the error correction for the errors self-correction
in the teleportation. For $|\phi_i\rangle ~(i=1,2,\cdots, 6)$, one
can recover the original state by performing generalized measurement
given by Kraus operators: \addtocounter{equation}{1}
\begin{align}
E'_{S1}&=| 0 \rangle\langle 0 |+\frac{a_0^3}{a_1^3}| 1 \rangle\langle 1 |+\frac{a_0^3}{a_2^3}| 2 \rangle\langle 2 |, \tag{\theequation a}\\
E'_{F1}&= \sqrt{1-\frac{a_0^6}{a_1^6}} | 1 \rangle\langle
1|+\sqrt{1-\frac{a_0^6}{a_2^6}} | 2 \rangle\langle 2 |
\tag{\theequation b}
\end{align}
 for $|\phi_1\rangle$ and \addtocounter{equation}{1}
\begin{align}
E'_{S2}&=\frac{a_0^3}{a_1^3}| 0 \rangle\langle 0 |+\frac{a_0^3}{a_2^3}| 1 \rangle\langle 1 |+| 2 \rangle\langle 2 |, \tag{\theequation a}\\
E'_{F2}&= \sqrt{1-\frac{a_0^6}{a_1^6}} | 0 \rangle\langle
0|+\sqrt{1-\frac{a_0^6}{a_2^6}} | 2 \rangle\langle 2 |
\tag{\theequation b}
\end{align}
 for $|\phi_2\rangle$ and \addtocounter{equation}{1}
\begin{align}
E'_{S3}&=\frac{a_0^3}{a_2^3}| 0 \rangle\langle 0 |+| 1 \rangle\langle 1 |+\frac{a_0^3}{a_1^3}| 2 \rangle\langle 2 |, \tag{\theequation a}\\
E'_{F3}&= \sqrt{1-\frac{a_0^6}{a_2^6}} | 0 \rangle\langle
0|+\sqrt{1-\frac{a_0^6}{a_1^6}} | 2 \rangle\langle 2 |
\tag{\theequation b}
\end{align}
 for $|\phi_3\rangle$ and \addtocounter{equation}{1}
\begin{align}
E'_{S4}&=| 0 \rangle\langle 0 |+\frac{a_0^2}{a_1a_2}| 1 \rangle\langle 1 |+\frac{a_0a_1}{a_2^2}| 2 \rangle\langle 2 |, \tag{\theequation a}\\
E'_{F4}&= \sqrt{1-\frac{a_0^4}{a_1^2a_2^2}} | 1 \rangle\langle
1|+\sqrt{1-\frac{a_0^2a_1^2}{a_2^4}} | 2 \rangle\langle 2 |
\tag{\theequation b}
\end{align}
 for $|\phi_4\rangle$ and \addtocounter{equation}{1}
\begin{align}
E'_{S5}&=\frac{a_0^2}{a_1a_2}| 0 \rangle\langle 0 |+\frac{a_0a_1}{a_2^2}| 1 \rangle\langle 1 |+| 2 \rangle\langle 2 |, \tag{\theequation a}\\
E'_{F5}&= \sqrt{1-\frac{a_0^4}{a_1^2a_2^2}} | 0 \rangle\langle
0|+\sqrt{1-\frac{a_0^2a_1^2}{a_2^4}} | 1 \rangle\langle 1 |
\tag{\theequation b}
\end{align}
 for $|\phi_5\rangle$ and \addtocounter{equation}{1}
\begin{align}
E'_{S6}&=\frac{a_0a_1}{a_2^2}| 0 \rangle\langle 0 |+| 1 \rangle\langle 1 |+\frac{a_0^2}{a_1a_2}| 2 \rangle\langle 2 |, \tag{\theequation a}\\
E'_{F6}&= \sqrt{1-\frac{a_0^2a_1^2}{a_2^4}} | 0 \rangle\langle 0|+
\sqrt{1-\frac{a_0^4}{a_1^2a_2^2}}| 2 \rangle\langle 2 |
\tag{\theequation b}
\end{align}
 for $|\phi_6\rangle$ respectively.

 For $|\phi_7\rangle, |\phi_8\rangle$ and $|\phi_9\rangle$, the
 recover operator are somewhat complicate because $a_0^2a_2 \leq
 a_1^2a_0$ for some entangled channels but $a_0^2a_2 \geq
 a_1^2a_0$ for other entangled channels. The  recover operators take different form
 for the above two cases.

 For the case of $a_0^2a_2 \leq  a_1^2a_0$, the recover Kraus
 operators take the form:
\addtocounter{equation}{1}
\begin{align}
E'_{S7}&=| 0 \rangle\langle 0 |+\frac{a_0a_2}{a_1^2}| 1 \rangle\langle 1 |+\frac{a_0^2}{a_1a_2}| 2 \rangle\langle 2 |, \tag{\theequation a}\\
E'_{F7}&=  \sqrt{1-\frac{a_0^2a_2^2}{a_1^4}}| 1 \rangle\langle
1|+\sqrt{1-\frac{a_0^4}{a_1^2a_2^2}} | 2 \rangle\langle 2 |
\tag{\theequation b}
\end{align}
 for $|\phi_7\rangle$ and
\addtocounter{equation}{1}
\begin{align}
E'_{S8}&=\frac{a_0a_2}{a_1^2}| 0 \rangle\langle 0 |+\frac{a_0^2}{a_1a_2}| 1 \rangle\langle 1 |+| 2 \rangle\langle 2 |, \tag{\theequation a}\\
E'_{F8}&=  \sqrt{1-\frac{a_0^2a_2^2}{a_1^4}}| 0 \rangle\langle
0|+\sqrt{1-\frac{a_0^4}{a_1^2a_2^2}} | 1 \rangle\langle 1 |
\tag{\theequation b}
\end{align}
for $|\phi_8\rangle$ and \addtocounter{equation}{1}
\begin{align}
E'_{S9}&=\frac{a_0^2}{a_1a_2}| 0 \rangle\langle 0 |+| 1 \rangle\langle 1 |+\frac{a_0a_2}{a_1^2}| 2 \rangle\langle 2 |, \tag{\theequation a}\\
E'_{F9}&= \sqrt{1-\frac{a_0^4}{a_1^2a_2^2}} | 0 \rangle\langle 0 |+
\sqrt{1-\frac{a_0^2a_2^2}{a_1^4}}| 2 \rangle\langle 2|
\tag{\theequation b}
\end{align}
for $|\phi_9\rangle$ respectively.

 On the other hand, if $a_0^2a_2 \geq
a_1^2a_0$, the recover Kraus
 operators take the form:
\addtocounter{equation}{1}
\begin{align}
E''_{S7}&=\frac{a_1^2}{a_0a_2}| 0 \rangle\langle 0 |+| 1 \rangle\langle 1 |+\frac{a_0a_1}{a_2^2}| 2 \rangle\langle 2 |, \tag{\theequation a}\\
E''_{F7}&= \sqrt{1-\frac{a_1^4}{a_1^0a_2^2}} | 0 \rangle\langle 0 |+
\sqrt{1-\frac{a_0^2a_1^2}{a_2^4}}| 2 \rangle\langle 2|
\tag{\theequation b}
\end{align}
for $|\phi_7\rangle$ \addtocounter{equation}{1}
\begin{align}
E''_{S8}&=| 0 \rangle\langle 0 |+\frac{a_0a_1}{a_2^2}| 1 \rangle\langle 1 |+\frac{a_1^2}{a_0a_2}| 2 \rangle\langle 2 |, \tag{\theequation a}\\
E''_{F8}&=  \sqrt{1-\frac{a_0^2a_1^2}{a_2^4}}| 1 \rangle\langle
1|+\sqrt{1-\frac{a_1^4}{a_0^2a_2^2}} | 2 \rangle\langle 2 |
\tag{\theequation b}
\end{align}
 for $|\phi_8\rangle$ and
\addtocounter{equation}{1}
\begin{align}
E''_{S9}&=\frac{a_0a_1}{a_2^2}| 0 \rangle\langle 0 |+\frac{a_1^2}{a_0a_2}| 1 \rangle\langle 1 |+| 2 \rangle\langle 2 |, \tag{\theequation a}\\
E''_{F9}&=  \sqrt{1-\frac{a_0^2a_1^2}{a_2^4}}| 0 \rangle\langle
0|+\sqrt{1-\frac{a_1^4}{a_0^2a_2^2}} | 1 \rangle\langle 1 |
\tag{\theequation b}
\end{align}
for $|\phi_9\rangle$ respectively.

 The probability of successfully
recovering the original state is
\begin{eqnarray}
 \nonumber P_G(4)&=& 3a_0^6+9a_0^4a_1^2+9\min\{a_0^4a_2^2,
 a_1^4a_0^2\}+6a_0^2a_1^2a_2^2.\\
\end{eqnarray}

We can easily see $P_G(4)\geq P_S(4)$ because of $a_2\geq a_1\geq
a_0$ .  It is obvious that for the maximally entangled channel, the
two protocols are equivalent, but for the non-maximally entangled
channel, GCTP is more efficient than SCTP. The total succuss
probability of SCTP is only determined by the smallest coefficient
of the entangled channel, on the other hand, the total succuss
probability of GCTP lies on all coefficients of the entangled
channel. For fixed $a_0$, the total succuss probability of GCTP
takes the minimum $P_G^{\rm min}(4)=6a_0^4+9a_0^6$  for the case of
$a_1=a_0$ and the maxmimum $P_G^{\rm
max}(4)=\frac{3}{2}a_0^2+6a_0^4-\frac{9}{2}a_0^6$ for the case of
$a_1=a_2$.

When one wants to apply global chain teleportation of a qutrit for
the case of arbitrary parties, a hindrance  occurs. With the
increase of the party number, the possible collapsing states of
receiver's qutrit will become numerous and disorderly. So the
error-correction operation of GCTP is very inconvenient. But we can
select a teleportation strategy eclectically between GCTP and SCTP.
Instead of being corrected by the final receiver in GCTP and
corrected by every parties in  SCTP, the errors can be corrected
every $n$ parties ($n$ is small). We call it piecewise globe chain
teleportation protocol (PGCTP). Here we select $n=3$ since error
self-correction only appears after making 3 times
 generalized  Bell-basis  measurements.

Suppose that Alice1 wants to teleport a quantum state $
|\psi\rangle=\alpha|0\rangle+\beta|1\rangle+\gamma|2\rangle$ to
Alice$3N+1$. There is no direct entangled resource between them, but
they can link through $3N-1$ intermediaries called Alice2, Alice3,
$\cdots$, Alice$3N$, respectively. Two neighboring parties share the
partially entangled state described by Eq.(2).

Firstly, Alice1, Alice2, Alice3  and Alice4 apply the GCTP according
to the steps as above described. After correct all errors, Alice4
obtain the quantum state $|\psi\rangle$ with the probability
$P_G(4)$. Next, Alice4, Alice5, Alice6 and Alice7 apply the GCTP
again. Thus Alice7 obtains the state $|\psi\rangle$ with the
probability $P_G(4)^2$, and so on. Finally, Alice$3N+1$ obtain the
quantum state $|\psi\rangle$ with the probability
\begin{equation}\label{pg}
   P_{\rm PG}(3N+1)=P_G(4)^N
\end{equation}
after correct the errors.
\begin{figure}[H]
  \centering
  \mbox{\subfigure[~$N=1$]{\includegraphics[width=1.6in]{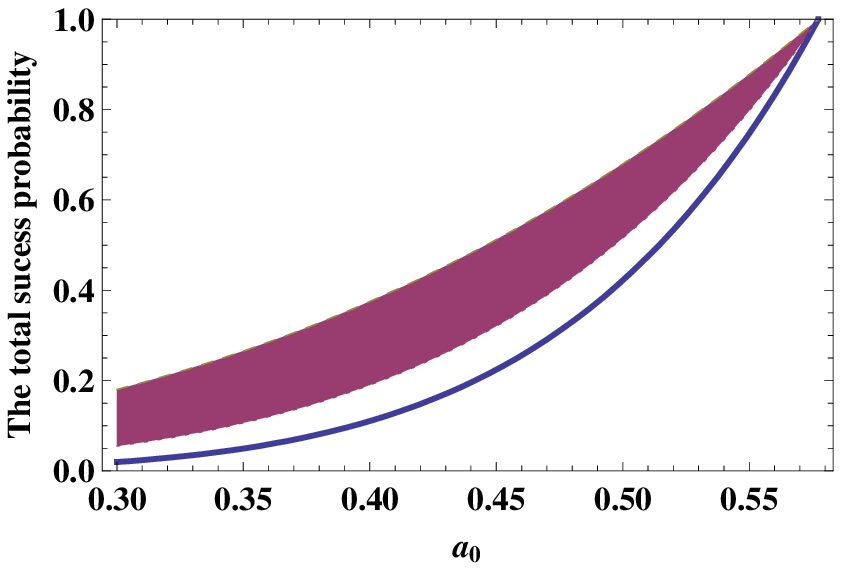}}\quad
\subfigure[~$N=2$]{\includegraphics[width=1.6in]{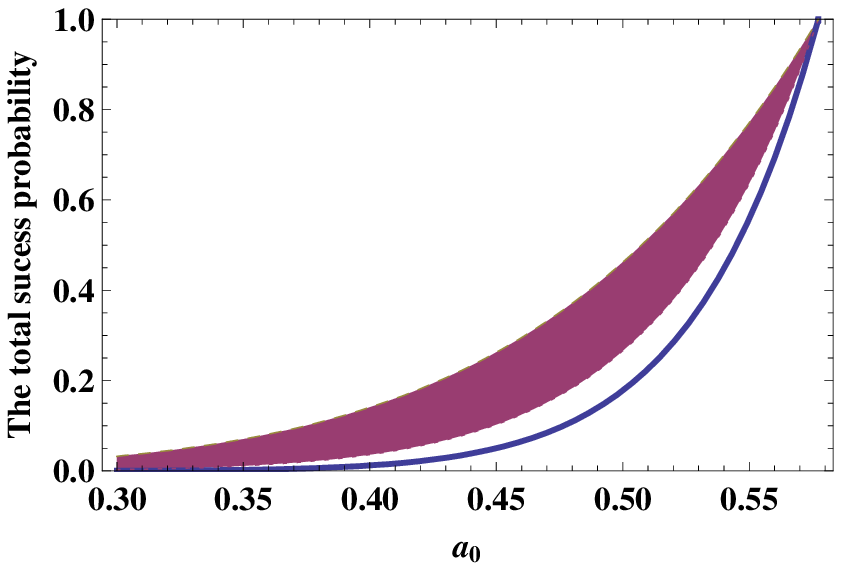}}}
\mbox{\subfigure[~$N=5$]{\includegraphics[width=1.6in]{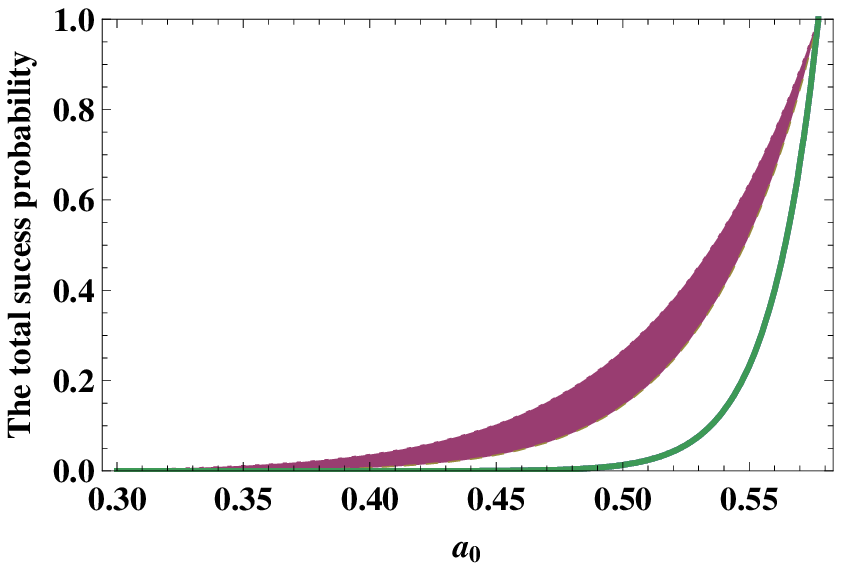}}\quad
\subfigure[~$N=10$]{\includegraphics[width=1.6in]{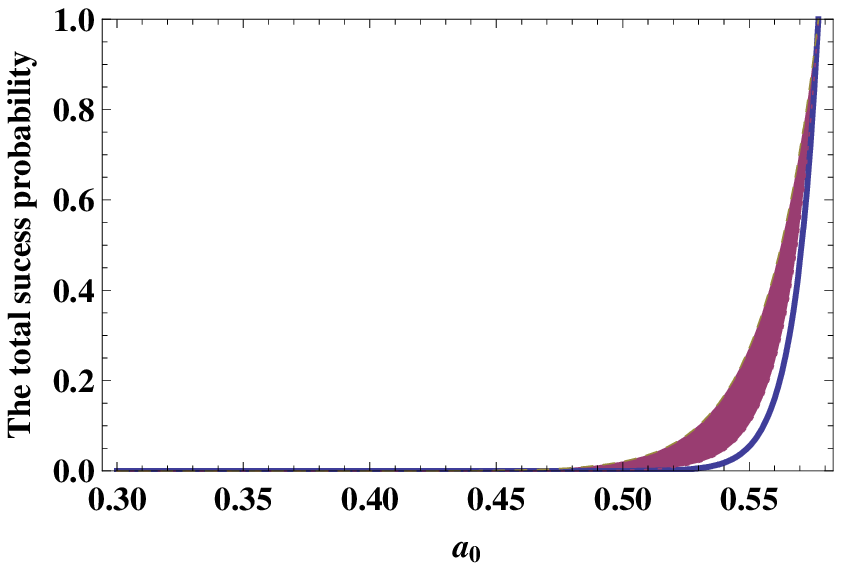}}}
\caption{The total success probability $P_{PG}$ and $P_S$ versus
$a_0$ for different $N$
   (Solid line: $P_S$, shadow region: $P_{\rm PG}$, where upper contour line corresponding to the case of $a_1=a_2$, and lower contour line corresponding to the case of $a_0=a_1$.).
   From (a) to(d), $N$ takes $1, 2, 5, 10$ accordingly.}
\end{figure}

In Fig.1, we plot $P_S$ and $P_{\rm PG}$ as the function of $a_0$
for different $N$. We can see that both the total success
probabilities of two protocols declines with the decrease of the
entanglement of channels. Moreover, the greater $N$ is, the more
sharper the success probabilities declines. It shows that the
quantum channel with small entanglement will become unpractical with
the increase of $N$.
 Fig.1 also indicates
explicitly that the  PGCTP is  more efficient than SCTP. For
example, for the case of $N=5$, the total success probability of
PGCTP $P_{\rm PG}^{\rm max}\approx 15\%$ while the total success
probability of SCTP $P_S$ only attains $1.3\%$ when the smallest
coefficient of channels $a_0=0.50$.
\newpage

The ratio of $P_{PG\rm}$ to $P_S$ as a function of $a_0$ for
different $N$ is illustrated in Fig.2. Here we only take $a_0$ from
$0.5$ to $\frac{1}{\sqrt{3}}$ because the small entanglement
channels are unpractical for large $N$.  From Fig.2, we can see that
the greater $N$ is, the larger $P_{\rm PG}/P_S$ is. In other words,
the efficiency of PGCTP is far higher than that of SCTP when the
steps of teleportation increase.

\begin{figure}
  \centering
  \mbox{\subfigure[~$N=1$]{\includegraphics[width=1.6in]{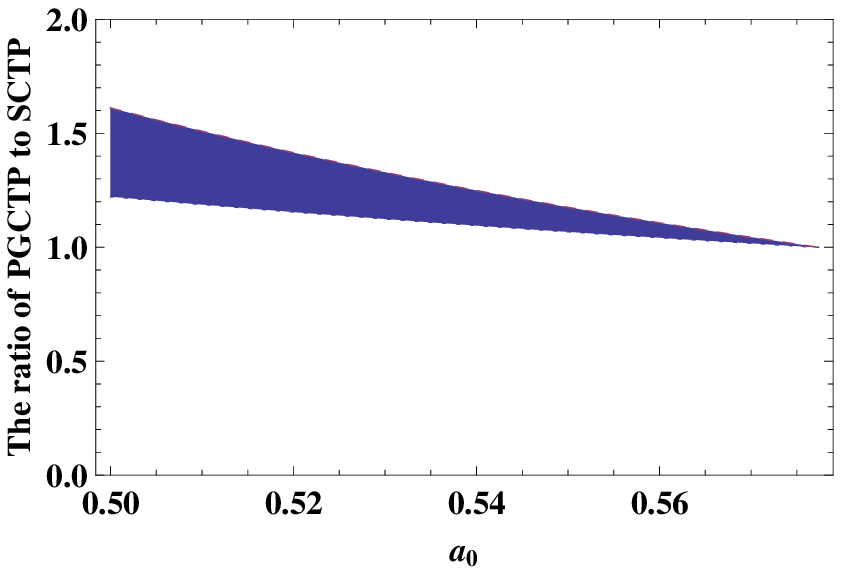}}\quad
\subfigure[~$N=2$]{\includegraphics[width=1.6in]{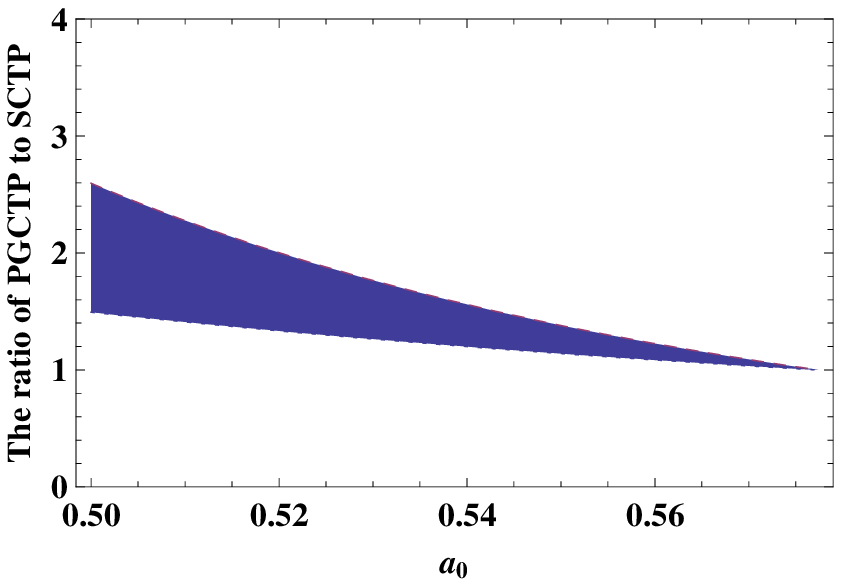}}}
\mbox{\subfigure[~$N=5$]{\includegraphics[width=1.6in]{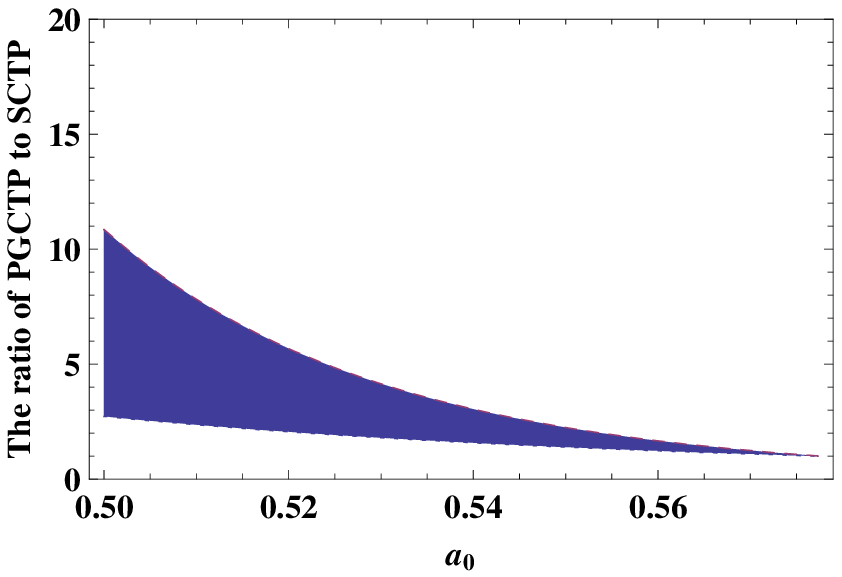}}\quad
\subfigure[~$N=10$]{\includegraphics[width=1.6in]{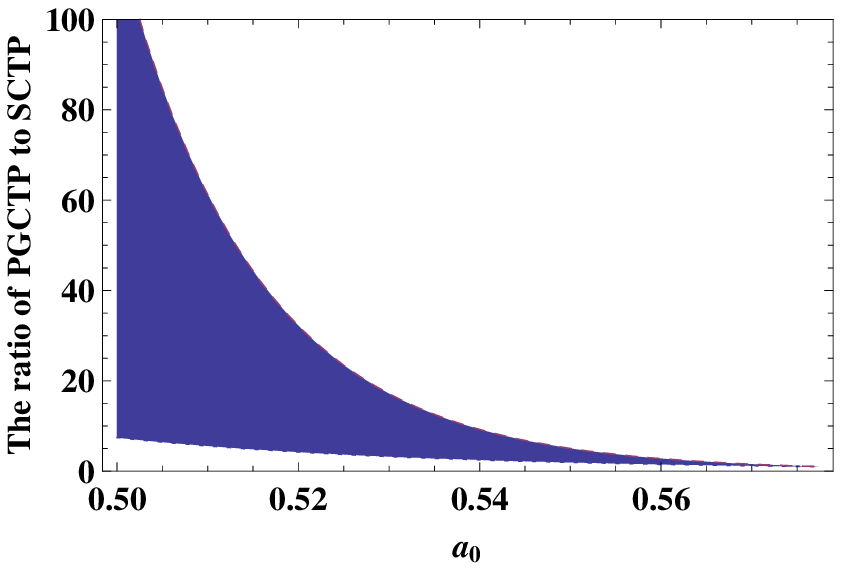}}}
\caption{The ratio of $P_{PG}$ to $P_S$ versus $a_0$ for different
$N$: upper contour line corresponding to the case of $a_1=a_2$, and
lower contour line corresponding to the case of $a_0=a_1$.  From (a)
to(d), $N$ takes $1, 2, 5, 10$ accordingly .}
\end{figure}

In summary, we have studied chain teleportation  of a qutrit-state
via the non-maximally two-qutrit entangled channels. For the case of
four parties, the efficiencies of two chain teleportation protocols,
the separate chain teleportation protocol and the global chain
teleportation protocol, are calculated. In SCTP the errors are
corrected between every step while in GCTP the errors are corrected
only at the end. With the increase of the parties number, possible
collapsing states of receiver's qutrit will become numerous and
disorderly. Therefore, the error-corrections of GCTP become very
inconvenient. We present a piecewise global chain teleportation
protocol for keeping away from this inconvenience. We show that
PGCTP is more efficient than SCTP.

\vspace{0.5cm}

{\noindent\bf Acknowledgments}\\[0.2cm]

This work was supported by the National Natural Science Foundation
of China under Grant No: 10971247, Hebei Natural Science Foundation
of China under Grant No: F2009000311 and  the Key Project of Science
and Technology Research of Education Ministry of China under Grant
No: 207011.


\end{document}